\documentclass[11pt,tightenlines]{revtex4}
\usepackage{hyperref}
\usepackage{amsfonts}
\usepackage{amsmath}
\usepackage{txfonts}
\usepackage{graphicx}
\usepackage{amssymb}
\usepackage{color}

\begin{document}
\title{Geometry of Spatial Memory Replay}
\author{Y. Dabaghian}
\address{Jan and Dan Duncan Neurological Research Institute, Baylor College of Medicine,
Houston, TX 77030, \\
and the Department of Computational and Applied Mathematics, 
Rice University, Houston, TX 77005\\
\\
e-mail: dabaghia@bcm.edu}

\date{\today}
\begin{abstract}
Place cells in the rat hippocampus play a key role in creating the animal's internal representation 
of the world. During active navigation, these cells spike only in discrete locations, together encoding 
a map of the environment. Electrophysiological recordings have shown that the animal can revisit 
this map mentally, during both sleep and awake states, reactivating the place cells that fired during 
its exploration in the same sequence they were originally activated. 
Although consistency of place cell activity during active navigation is arguably enforced by sensory and 
proprioceptive inputs, it remains unclear how a consistent representation of space can be maintained 
during spontaneous replay. We propose a model that can account for this phenomenon and suggests 
that a spatially consistent replay requires a number of constraints on the hippocampal network that 
affect its synaptic architecture and the statistics of synaptic connection strengths.
\end{abstract}
\maketitle

\newpage

\section{Introduction}
\label{section:intro}

\begin{figure}[ht]
\hfill
\begin{center}
\includegraphics[scale=0.75]{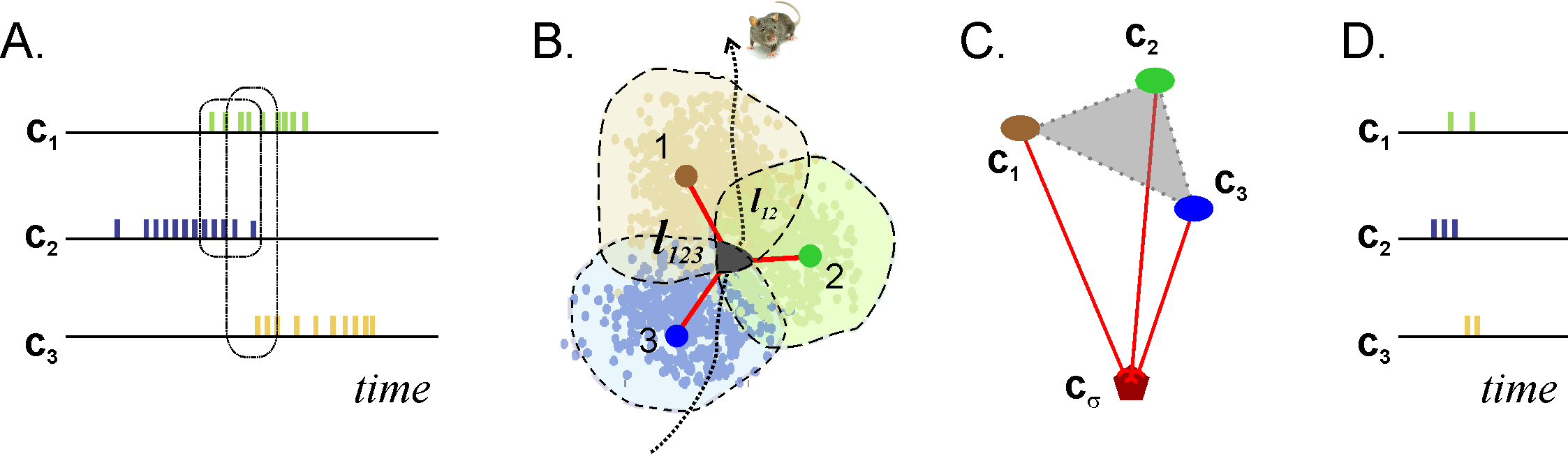}\end{center}
\caption{{\bf Place cells, place fields and cell assemblies}. 
{\bf A.} A schematic representation of the spike trains produced by three place cells, $c_1$, $c_2$ and $c_3$. 
The two red rectangles mark the periods during which the cells are coactive \cite{Arai}. {\bf B.} The gold, 
green and blue areas represent place fields. Place cell firing rate is maximal at the center of the place field
and attenuates towards its periphery; this pattern can be closely approximated by Gaussian distribution. 
Place cell cofiring reflects overlap between respective place fields: cells $c_1$ and $c_2$ are coactive in the 
location $l_{12}$, cells $c_2$ and $c_3$ are coactive in the location $l_{23}$ and so on. The red links mark 
distances between the centers of the place fields and the triple overlap domain,  $l_{123}$ (dark region in the 
center). {\bf C.} A schematic representation of a cell assembly: the three place cells on the top synapse onto 
a readout neuron (red dot), which activates within the cell assembly field $l_{123}$. {\bf D.} During replay, 
the place cells repeat on a millisecond time scale the order of spiking that they exhibit during active navigation.}
\label{PCs}
\end{figure}

In the course of learning a spatial environment, an animal forms an internal representation of space 
that enables spatial navigation and planning \cite{Schmidt}. The hippocampus plays a key role 
in producing this map through the activity of location-specific place cells \cite{OKeefe}. At the 
neurophysiological level, these place cells exhibit spatially selective spiking activity. As the animal 
navigates its environment, the place cell fires only at a discrete location---its place field (Figure~\ref{PCs}A-B). 
It is believed that the entire ensemble of place cells serves as a neuronal basis of the animal's spatial 
awareness \cite{McNaughton1,Best}.

Remarkably, place cells spike not only during active navigation but also during quiescent wake states 
\cite{Pfeiffer,Davidson} and even during sleep \cite{Louie,Skaggs,Wilson1}. For example, the animal 
can ``replay'' place cells in sequences that correspond to the physical routes traversed during active 
navigation \cite{Foster,Diba,Hasselmo} or ``preplay'' sequences that represent possible future 
trajectories, either in direct or reversed order, while pausing at a decision point \cite{Johnson,Pastalkova}. 
This phenomenon implies that, after learning, the animal can explore and retrieve spatial information by 
cuing the hippocampal network \cite{MoserReplay,Dragoi}, which may in turn be viewed as a 
physiological correlate of ``mental exploration'' \cite{Hopfield,Hasselmo1}.

It bears noting, however, that the actual functional units for spatial information processing in the 
hippocampal network are not individual cells but repeatedly activated groups of place cells known as 
cell assemblies (see \cite{Buzsaki} and Figure~\ref{PCs}C). Although the physiological properties of 
the place cell assemblies remain largely unknown, it is believed that the cells constituting an assembly 
synaptically drive a certain readout unit downstream from the hippocampus. In the ``reader-centric'' 
view, this readout neuron---a small network or, most likely, a single neuron---is what actually defines
the cell assembly, by actualizing the information provided by its activity \cite{Buzsaki}.
The identity of the readout neurons in some cases is suggested by the network's anatomy. For example, 
there are direct many-to-one projections from the CA3 region of the hippocampus to the CA1 region. 
Since replays are believed to be initiated in CA3 \cite{Carr,Johnson}, this implies that the CA1 place cells 
may serve as the readout neurons for the activity of the CA3 place cells. 
Assuming that contemporaneous spiking of place cells implies overlap of their respective place fields 
(Figure~\ref{PCs}A-B), it is possible to decode the rat's current location from the ongoing spiking activity 
of a mere 40-50 neurons \cite{Brown}. This suggests that the readout neurons may be wired to encode 
spatial connectivity between place fields by responding to place cell coactivity (see Figure~\ref{PCs}A-C 
and \cite{Jarsky,Katz,eLife}).

A natural assumption underlying both the trajectory reconstructing algorithms \cite{Brown} and various 
path integration models \cite{McNaughton1,Samsonovich,Issa} is that the representation of spatial 
locations during physical navigation is reproducible. If the rat begins locomotion at a certain location and 
at a certain moment of time, $t_0$, and then returns to the same location at a later time, $t_1$, then the 
population activity of the place cells at $t_0$ and $t_1$ is the same. Similarly, if spatial information is 
consistently represented during replays, then the activity packet in the hippocampal network should be 
restored upon ``replaying" a closed path. Whereas the correspondence between place cell activity and 
spatial locations (i.e., place fields) during physical navigation is enforced by sensory and proprioceptive 
inputs \cite{Samsonovich}, the consistency of spatial representation during replay must be attributable 
solely to the network's internal dynamics \cite{Gupta1}.

Here we develop a model that accounts for how a neuronal network could maintain consistency of spatial 
information over the course of multiple replays or preplays. This model is based on the discrete differential 
geometry theory developed in \cite{Novikov}, which reveals that key geometric concepts can be expressed 
in purely combinatoric terms. The choice of this theory is driven in part by recent work that indicates that the 
hippocampus provides a topological framework for spatial information rather than a geometric or Cartesian 
map \cite{eLife,Alvernhe,Wu}.

The results suggest that to maintain consistency of spatial information during  path replay, the synaptic 
connections between the place cells and the readout neurons must adhere to a zero holonomy principle.

\section{The Model}
\label{section:model}

{\bf The simplicial model of the cell assembly network}.
A convenient framework for representing a population of place cell assemblies is provided by 
simplicial topology \cite{Bredon,Dubrovnik,Alexandrov}. In this approach, an assembly of $d+1$ 
place cells, $\{c_{i_0}$, $c_{i_1}$, ..., $c_{i_d}\}$, is represented by a $d-$dimensional abstract 
simplex (not to be confused with a geometric simplex) containing $d+1$ vertexes, $\sigma=\left[ 
v_{i_0},v_{i_1},...,v_{i_d}\right]$, where each vertex, $v_i$, corresponds to a place cell $c_i$ 
(in the following, the same symbol ``$\sigma$'' will be used to denote a cell assembly and the 
simplex that represents it)  \cite{PLOS,Arai}. 
The entire network can then be represented by a purely combinatorial simplicial complex $\mathcal{T}$ 
\cite{Alexandrov,Prasolov} whose \emph{maximal} simplexes correspond to place cell assemblies 
\cite{Comb}. Simplexes in $\mathcal{T}$ may overlap: physiological studies demonstrate that 
a given place cell may be a part of many cell assemblies \cite{Georgopoulos,Tudusciuc}. 
Many authors have suggested that place cell assemblies should overlap significantly in order to better 
represent contiguous spatial locations \cite{Curto,Jahans,Maurer,Gupta}: the more cells shared by 
$\sigma_1$ and $\sigma_2$, the closer the encoded locations are to one another. The most detailed 
representation of the environment is produced by a population of maximally overlapping cell assemblies, 
which differ by a single cell. In such case, a transition of the activity from one cell assembly $\sigma_1$ 
to another $\sigma_2$ occurs when one place cell in $\sigma_1$ turns off and another cell in the new 
assembly $\sigma_{2}$ turns on. The resulting simplicial complex $\mathcal{T}$ has the structure of a 
combinatorial $d$-dimensional simplicial manifold (in the literature also referred to as ``pure complex'' 
or ``pseudomanifold'', \cite{Prasolov}).

{\bf Population activity in the cell assembly complex $\mathcal{T}$}. The simplicial complex 
$\mathcal{T}$ is a convenient instrument for relating place cell coactivity to the topology of 
the rat's environment \cite{PLOS,Arai}. The rat's movements in the physical environment 
induce a packet of place cell activity that propagates in 
the hippocampal network---an ``activity bump'' \cite{Samsonovich}. In our model 
the propagation of the activity bump corresponds to an ``active simplex'' propagating 
through $\mathcal{T}$. The resulting population activity vector is then
\begin{equation}
\mathbf{f}_{\sigma}^{\top}=(f_{\sigma},f_{\sigma,v_{i_0}},
f_{\sigma,v_{i_1}}, ...,f_{\sigma,v_{i_k}}),
\label{act}
\end{equation}
where the first component $f_{\sigma}$ represents the spiking rate of the readout 
neuron, and $f_{\sigma,v_i}$ denotes the spiking rate of the place cell $c_i$ within 
the assembly $\sigma$. Roughly speaking, $f_{\sigma,v_i}$ can be viewed as the 
firing rate of $c_i$ at the location where the place fields of the cells constituting the 
assembly $\sigma$ overlap, which we refer to as the cell assembly field, $l_{\sigma}$ 
(the domain $l_{123}$ on Figure~\ref{PCs}B). 
A given place cell $c_i$ is a part of many cell assemblies $\sigma_1, \sigma_2, ...$, 
whose fields $l_{\sigma_1}, l_{\sigma_2}, ...$ are contained in the $c_i$'s place field;
thus, the higher the orders of the cell assemblies, the (statistically) smaller the 
$l_{\sigma}$s \cite{Comb}.
Since the individual place cell spiking rates are well approximated by smooth Gaussian 
functions of the rat's coordinates \cite{Eden}, the quantities $f_{\sigma,v_i}$ remain 
approximately constant over $l_{\sigma}$.  
The components of the population activity vector (\ref{act}) in a given cell assembly 
can then be related to the corresponding place cells' maximal firing rates $f_{v_i}$ 
by a set of factors $0\leq h_{\sigma,v_i}\leq 1$, that are specific to a given cell and 
a given cell assembly, 
\begin{equation}
f_{\sigma,v_i}=h_{\sigma,v_i}\,f_{v_i},
\label{fsig}
\end{equation}
which may be viewed as measures of the separation between the location $l_{\sigma}$ 
and the respective place fields' centers (Figure~\ref{PCs}B). In other words, the coefficients 
$h_{\sigma,v_i}$ provide a discrete description of the place field map's geometry.

As the rat moves from one cell assembly field to another (e.g., from $l_{\sigma_1}$ 
to $l_{\sigma_2}$), the activity packet in $\mathcal{T}$ shifts from the maximal simplex 
$\sigma_1$ to the maximal simplex $\sigma_2$, then to $\sigma_{3}$, and so on, tracing 
a ``simplicial path,'' 
\begin{equation}
\Gamma =\left\langle \sigma_1,\sigma_2,...,\sigma_{n}\right\rangle,
\label{G}
\end{equation}
(these are ``thick paths" in the terminology of \cite{Novikov}). As a result, every path 
$\gamma $ in the rat's physical environment 
corresponds to a simplicial path $\Gamma\in\mathcal{T}$, which can be viewed as an 
abstract representation of the place cell trajectory code used in \cite{Brown}. However, 
in order to represent the path $\Gamma$ in the hippocampal network, the activity of 
each place cell assembly $\sigma\in\Gamma$ should activate the corresponding readout 
neuron $c_{\sigma}$. Thus, during the activation period, the net input from the presynaptic 
cells in $\sigma$ should exceed the corresponding readout neuron's firing threshold 
$\theta_{\sigma}$, 
\begin{equation}
\sum_{v_i\in \sigma}w_{\sigma,v_i}f_{\sigma,v_i}\geq\theta_{\sigma},
\label{wf}
\end{equation}
where the coefficients $w_{\sigma,v_i}$ represent the strengths of synaptic connections 
between the place cells and the readout neuron \cite{Buzsaki,Legenstein}.  

In other words, this is a rate model in that the activity of cells is described by a single parameter: 
the firing rate, $f$, related via coefficients $h$ to the maximal rate (\ref{fsig}). If the network is trained
---the synaptic architecture is fixed, place fields are stable---then each cell assembly fires when the 
rat visits (or replays) a specific spot where the respective place fields overlap. Because this spot is 
very small compared to the size of place fields, the left side of (\ref{wf}) is the essentially the same 
every time. 

Using (\ref{fsig}), the condition (\ref{wf}) becomes
\begin{equation}
\sum_{v_i\in \sigma}b_{\sigma,v_i}f_{v_i}\geq\theta_{\sigma},
\label{th}
\end{equation}
where the variables $\theta$ and
\begin{equation}
b_{\sigma,v_i}=w_{\sigma,v_i}h_{\sigma,v_i},
\label{b}
\end{equation}
are defined on all simplexes (i.e., all cell assemblies), and the variable $f$ on the 
 vertexes (i.e., place cell). 

{\bf Dressed cell assembly complex $\mathcal{T}_{\ast}$}. The coefficients $b_{\sigma,v_i}$ can 
be regarded as characteristics of the maximal simplexes of $\mathcal{T}$ and the values $f_{v_i}$ 
as characteristics of its vertexes. Together, these parameters produce a ``dressing" of the cell 
assembly complex with physiological information about the cells' spiking and the network's synaptic 
architecture. Equation (\ref{th}) singles out a set $\mathcal{B}_{\mathcal{T}_{\ast}}$ of \emph{valid} 
dressings, $b_{\mathcal{T}_{\ast}}=\{b_{\sigma,v}: \mathrm{eq.\, (\ref{th}) \,\,is\,\,satisfied}\}$ 
which enable readout neurons to respond to presynaptic activity and thus defines the scope of 
working synaptic architectures of the place cell assembly networks.

Replays occur on a millisecond time scale and produce only a few spikes per activity period \cite{Hasselmo1} 
(Figure~\ref{PCs}D), which is comparable to the stochastic background activity of neurons. In order to 
distinguish cell assembly activation from the assembly's background activity, the readout neuron should be 
tuned to a particular combination of inputs; the physiological process most likely involves gating 
specific parts of the dendritic tree by timed inputs from the presynaptic cells. Here, the model 
employs a simplified version of this process: we propose that the readout neuron should remain 
in a sensitive, near-threshold state \cite{Thresh} that allows it to quickly respond to the cell assembly 
during each individual step of replay. Thus, we hypothesize that during replays the inequality (\ref{th}) may 
be approximated by the equation 
\begin{equation}
\sum_{v_i\in \sigma}b_{\sigma,v_i}f_{v_i}=\theta_{\sigma},
\label{Q}
\end{equation}
which further restricts the set of valid dressings to a special marginal set $\bar {\mathcal B}_{{\mathcal T}_{\ast}}$, 
for which spontaneous replays of place cell assemblies are possible. Note, however, that equation (\ref{Q}) does not 
fix the values of the parameters $f_{v_i}$ and $b_{\sigma,v_i}$, and so it allows a significant variability of spiking 
activity and of the synaptic connection strengths. To emphasize the assumption that place cell activity during replays 
elicits a response from the readout neuron at a constant rate $f_{\sigma}>0$, it is convenient to use the notation 
\begin{equation}
\theta_{\sigma}\equiv b_{\sigma,\sigma}f_{\sigma},
\label{thf}
\end{equation}
where $b_{\sigma,\sigma}>0$ is a fixed parameter that can be interpreted as the 
readout neuron's susceptibility to discharge. 
\begin{figure} 
\includegraphics[scale=0.65]{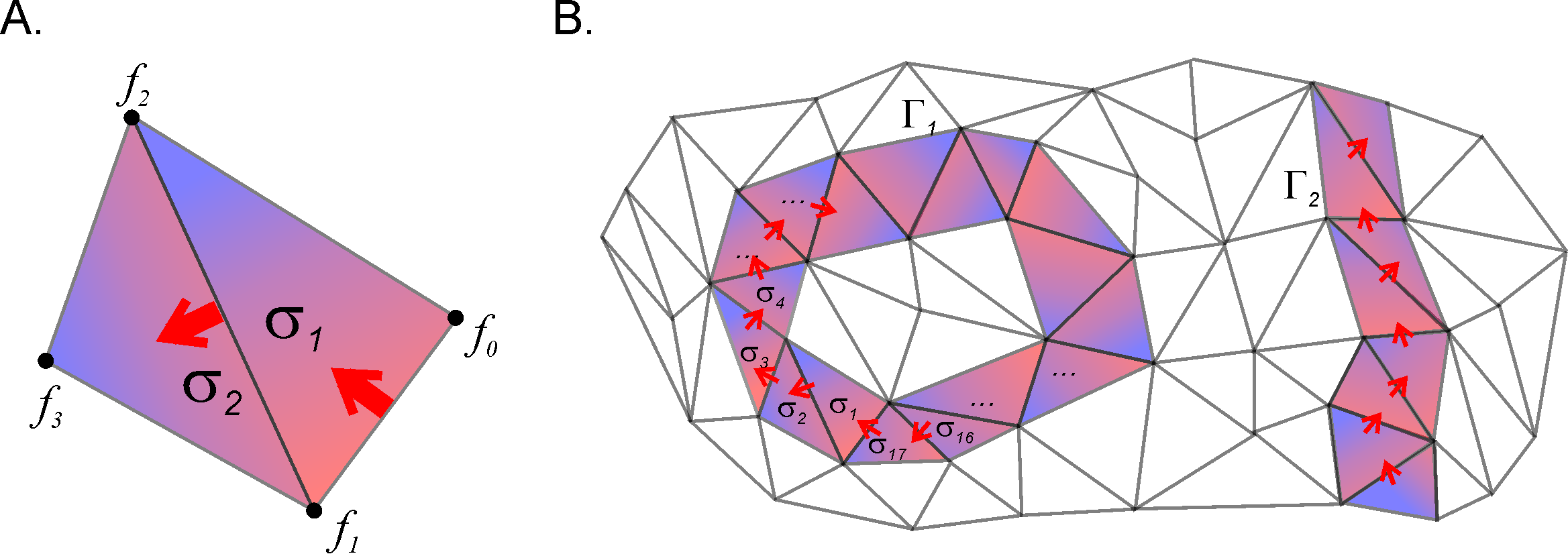}
\caption{\label{ThickPath} {\bf Propagation of place cell activity along simplicial 
paths}. {\bf A.} The firing rate of the cell $c_2$, required to activate the readout 
neuron of the cell assembly $\sigma_1$, can be inferred from the firing rates of 
the other two cells, $c_0$ and $c_1$ using the equation (\ref{v2eq}) (red arrow 
over $\sigma_1$). The the resulting rate $f_2$ of $c_2$ and the rate $f_0$ of 
$c_0$ define the rate $f_3$ of the cell $c_3$ required to activate the readout 
neuron in the next cell assembly $\sigma_2$ according to the equation (\ref{v2sol1}) 
(red arrow over the $\sigma_{2}$). {\bf B.} A network of maximally overlapping cell 
assemblies is represented by a simplicial manifold $\mathcal{T}_{\ast}$. The 
replayed sequences correspond to simplicial paths that can be closed ($\Gamma_1$) 
or open ($\Gamma_2$).}
\end{figure} 

{\bf Replays}.
Equation (\ref{Q}) defined at each simplex of $\mathcal{T}_{\ast}$ \cite{Novikov} 
provides a simple tool for building a model of hippocampal replay. As an illustration, 
consider the case when $\mathcal{T}_{\ast}$ is two-dimensional and let 
$\sigma_1=\left[ v_0,v_1,v_2\right]$ be a $2D$ simplex representing an assembly 
of three cells with the firing rates $f_{v_0}$, $f_{v_1}$ and $f_{v_2}$. If equation 
(\ref{Q}) holds over $\sigma_1$, then the readout neuron $c_{\sigma_1}$ fires with 
the rate $f_{\sigma_1}$ in response to the coactivity of $c_1$, $c_2$ and $c_3$,
\begin{equation}
b_{\sigma_1,v_0}f_{v_0}+b_{\sigma_1,v_1}f_{v_1}+b_{\sigma_1,v_2}f_{v_2}
=b_{\sigma_1,\sigma_1}f_{\sigma_1}.
\label{v1eq}
\end{equation}
Suppose that equation (\ref{Q}) also holds for an adjacent (maximally overlapping) 
cell assembly, represented by an adjacent simplex $\sigma_2=\left[ v_1,v_2,v_3\right]$, 
so that the second readout neuron $c_{\sigma_2}$ fires with the rate $f_{\sigma_2}$
\begin{equation}
b_{\sigma_2,v_1}f_{v_1}+b_{\sigma_2,v_2}f_{v_2}+b_{\sigma_2,v_3}f_{v_3}
=b_{\sigma_2,\sigma_2}f_{\sigma_2}.
\label{v2eq}
\end{equation}
A key observation here is that, since $\sigma_2$ shares vertexes $v_1$ and $v_2$ with 
$\sigma_1$, the corresponding firing rates $f_{v_1}$ and $f_{v_2}$ in (\ref{v2eq}) 
define uniquely the firing rate of the remaining cell, $f_{v_3}$, required to activate the readout 
neuron $c_{\sigma_2}$ (Fig.~\ref{ThickPath}A),
\begin{equation}
f_{v_3}=\frac{1}{b_{\sigma_2,v_3}}(b_{\sigma_2,\sigma_2}
f_{\sigma_2}-b_{\sigma_2,v_1}f_{v_1}-b_{\sigma_2,v_2}f_{v_2}).
\label{v2sol1}
\end{equation}
Similarly, if there is another $2D$ simplex $\sigma_3=[ v_2,v_3,v_4]$ adjacent to 
$\sigma_2$, then, once the value $f_{v_3}$ is found from (\ref{v2sol1}), the firing 
rate at $v_4$ can be obtained from $f_{v_2}$ and $f_{v_3}$, and so on 
(Figure~\ref{ThickPath}B). 
In other words, once the synaptic connections $b_{\sigma,v}$ are specified for all simplexes, 
equation (\ref{Q}) can be used to describe the conditions for transferring the activity vector 
$\mathbf{f}_{\sigma}$ over the entire complex $\mathcal{T}_{\ast}$ \cite{Novikov}. Notice 
however, that equations (\ref{v1eq})-(\ref{v2sol1}) do not specify the mechanism responsible 
for generating place cell activity; they only describe the conditions required to ignite the cell 
assemblies in a particular sequence. 
While the subsequent simplexes $\sigma_n$ and $\sigma_{n+1}$ in the simplicial path (\ref{G}) 
are not necessarily adjacent, the activity according to equation (\ref{Q}) is propagated along a 
sequence of adjacent maximal simplexes, such as depicted in Figure~\ref{ThickPath}B.

{\bf Discrete Holonomy}. Using the notation 
\begin{equation}
\mu_{x,y}^{\sigma}\equiv \frac{b_{\sigma,x}}{b_{\sigma,y}},
\label{mu}
\end{equation}
equation (\ref{v2sol1}) defined over a simplex $\sigma_p$ can be rewritten in matrix form
\begin{equation}
\mathbf{f}_{q}=M_{q,p}^{v_t v_s}\mathbf{f}_{p},
\label{prop}
\end{equation}
where the ``transfer matrix'' $M_{q,p}^{v_t v_s}$ propagates the population activity vector 
from the incoming facet of the simplex $\sigma_p$ into the activity vector of the outgoing, 
opposite facet shared with the next simplex $\sigma_q$ (edges  $[v_0, v_1]$ and $[v_1,v_2]$ 
respectively on Figure~\ref{ThickPath}A), in which the vertex $v_{s}$ of the simplex $\sigma_p$ 
shuts off and the vertex $v_{t}$ of the adjacent simplex activates.
If there is a total of $n$ simplexes in the path $\Gamma $ ($n=17$ for the closed simplicial path
$\Gamma_{1}$ shown on Figure~\ref{ThickPath}B) then the corresponding chain of $n$ equations 
(\ref{prop}) will produce
\begin{equation}
\mathbf{f}_{n}=
M_{q_n,p_n}^{v_{t_n}v_{s_n}}M_{q_{n-1},p_{n-1}}^{v_{t_{n-1}}v_{s_{n-1}}}
...M_{q_1,p_1}^{v_{t_1}v_{s_1}}\mathbf{f}_{1}.
\label{kk}
\end{equation}

If the simplicial path is closed, then the activity vector should be restored upon completing the 
loop, i.e., $\mathbf{f}_{n}=\mathbf{f}_1$. According to (\ref{kk}), this will happen if the product 
of the transfer matrices along $\Gamma$ yields a unit matrix,
\begin{equation}
M_{\Gamma }\equiv 
M_{q_n,p_n}^{v_{t_n}v_{s_n}}M_{q_{n-1},p_{n-1}}^{v_{t_{n-1}}v_{s_{n-1}}}
...M_{q_1,p_1}^{v_{t_1}v_{s_1}}=\mathbf{1}.
\label{prod1}
\end{equation}
It can be directly verified, however, that condition (\ref{prod1}) is not satisfied automatically: the 
product of transfer matrices (\ref{prod1}) has the structure
\begin{equation}
M_{\Gamma }=\left( 
\begin{array}{ccc}
1 & 0 & 0 \\ 
0 & 1 & 0 \\ 
\kappa_{\Gamma,1} & \kappa_{\Gamma,2} & 1+\kappa_{\Gamma,3}
\end{array}
\right),
\label{MGamma}
\end{equation}
which differs from the unit matrix $M_{\Gamma }\neq \mathbf{1}$ (see Appendix). This implies that 
a population activity vector $\mathbf{f}_{\sigma}$ is in general altered by translations around closed 
simplicial paths, $\mathbf{f}_{\sigma}(t_{\rm start}) \neq\mathbf{f}_{\sigma}(t_{\rm end})$.
To formulate this another way, the spiking condition (\ref{Q}) does not automatically guarantee that 
the readout neurons will consistently represent spatial connectivity; the latter requires additional 
constraints (\ref{prod1}), irrespective of the mechanism that shifts the activity bump. 
\begin{figure} 
\includegraphics[scale=0.7]{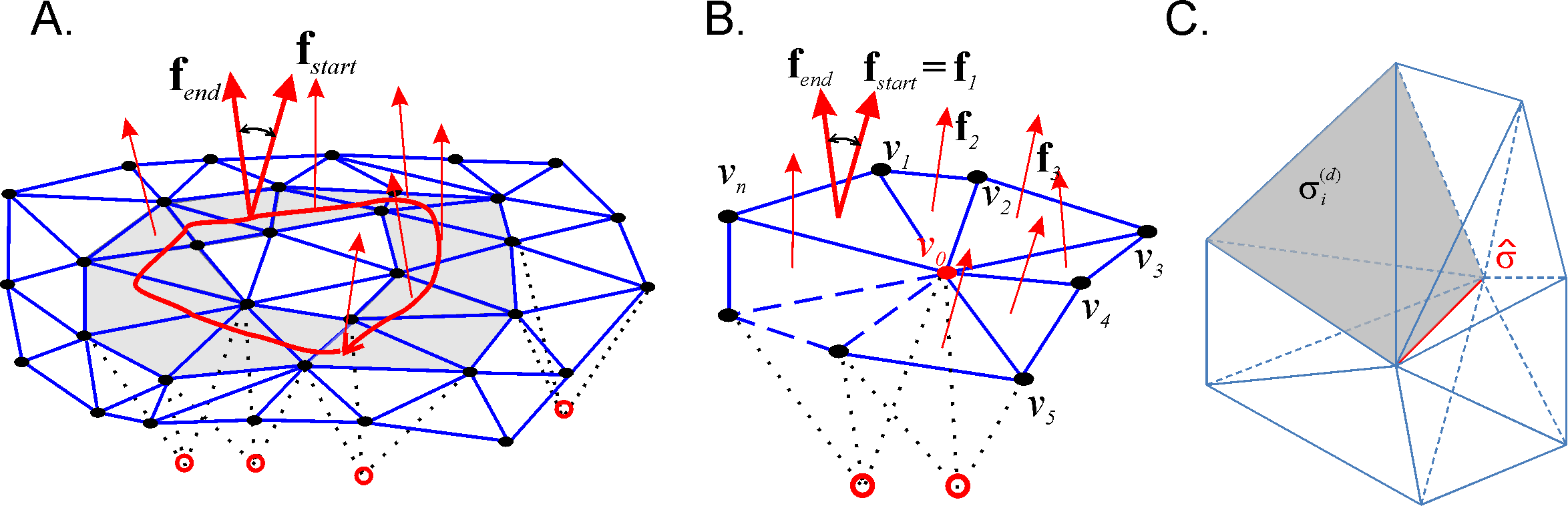}
\caption{\label{Holonomy} {\bf Discrete geometry of a dressed simplicial complex}. 
{\bf A.} Discrete holonomy: a population activity vector (red arrow) changes its direction from simplex to 
simplex as described by (\ref{prop}). Upon completing a closed path, the starting and ending vectors may 
differ, $\mathbf{f}_{start}\neq\mathbf{f}_{end}$, which indicates nonzero holonomy. {\bf B.} A $2D$ 
elementary closed path of the order $n$ encircling a vertex $v_0$. The ``pivot" vertex $v_0$ carries the 
discrete curvature coefficients defined by (\ref{curv}). {\bf C.} A higher dimensional elementary closed 
path consisting of $d$-dimensional simplexes (one such exemplary simplex $\sigma_1^{(d)}$ is shadowed) 
sharing the same $(d-2)$-dimensional face, the pivot simplex $\hat\sigma_1$, shown in red. The $d-2$ 
dimensional pivot simplex $\hat \sigma$ shown in red carries the curvature 
coefficients $\kappa_{\hat \sigma,i}$.}
\end{figure} 

Mathematically, a mismatch between the starting and the ending orientation of the population activity 
vector is akin to the differential-geometric notion of holonomy which, on Riemannian manifolds, 
measures the change of a vector's orientation as a result of a parallel transport around a closed loop 
\cite{Novikov,Bredon,Dubrovnik,Sternberg}. Hence, the requirement (\ref{prod1}) that the activity 
vector should be the same after completing a closed simplicial trajectory implies that the discrete 
holonomy along paths in $\mathcal{T}_{\ast}$ should vanish.

{\bf Discrete Curvature}.
In differential geometry, zero holonomy on a Riemannian manifold is achieved by
requiring that the Riemannian curvature tensor $R^{i}_{jkl}$ associated with the
connection $\Gamma^{i}_{jk}$ vanishes at every point $x$ \cite{Dubrovnik,Sternberg}. 
This condition is established by contracting closed paths to infinitesimally small 
loops encircling a point $x$ and translating in parallel a unit vector $\vec n$ around 
that loop. The difference between the starting and the ending orientations of $\vec n$
defines the curvature at the point $x$ \cite{Sternberg}. An analogous procedure can be 
performed on a discrete manifold $\mathcal{T}_{\ast}$. However, there is a natural limit 
to shrinking simplicial paths: in a $d$-dimensional complex, the tightest simplicial paths 
consist of $d-$dimensional simplexes $\sigma^{(d)}$ which intersect the same $(d-2)$ 
dimensional face (see Figure~\ref{Holonomy}B). Such a path $\Gamma_{\sigma^{(d-2)}}$ 
we will call an ``elementary closed path'', following \cite{Novikov}. The order $s_{n}$ 
of such a path is defined by the number of $d$-dimensional simplexes $\sigma^{(d)}$ 
encircling a simplex $\sigma^{(d-2)}$. In the following we will use the short notation 
$\hat \sigma$ for the ``pivot" simplexes $\hat\sigma\equiv\sigma^{(d-2)}$ whereas the 
elementary simplicial path encircling $\hat\sigma$ will be denoted as $\Gamma_{\hat \sigma}$.

In order to ensure zero holonomy of place cell activity along \emph{all} closed paths in 
$\mathcal{T}_{\ast}$, it is sufficient to verify that the holonomy vanishes for all elementary 
closed paths \cite{Novikov}. The product of the matrices $M_{q,p}^{v_t v_s}$ encircling the 
pivot $\hat \sigma$ (Figure~\ref{Holonomy}B) has the same form as equation (\ref{MGamma});
however, the coefficients $\kappa_{\hat\sigma,i}$ at the bottom row of the matrix $M_{\hat\sigma}$ 
can be viewed as the curvatures defined at $\hat\sigma$. Thus, to ensure zero holonomies, the 
conditions
\begin{equation}
\kappa_{\hat\sigma,i}=0, 
\label{curv}
\end{equation}
$i=1$, ..., $d+1$, must be imposed on the connection coefficients $b_{\sigma,v_i}$ 
at every pivot simplex $\hat\sigma$ of a $d$-dimensional dressed cell assembly 
simplicial complex. 
For example, an elementary $2D$ closed path encircling a vertex $v_{0}$ with $n$ 
simplexes enumerated as shown on Figure~\ref{Holonomy}C yields the holonomy matrix
$$M_{v_{0}}=\left( 
\begin{array}{ccc}
1 & 0 & 0 \\ 
0 & 1 & 0 \\ 
\kappa_{v_{0},1} & \kappa_{v_{0},2} & 1+\kappa_{v_{0},3}
\end{array}\right). $$ 
The values $\kappa_{v_{0},i}$, $i=1$, $2$, $3$, of the bottom row that distinguish 
$M_{v_0}$ from the unit matrix should be considered as discrete curvatures defined 
at the pivot vertex $v_{0}$ (see Figure~\ref{Holonomy}C and \cite{Novikov}), which 
need to vanish in order to ensure a consistent representation of space during replays. 

Since there exists a finite number of pivot simplexes, the number of constraints (\ref{curv}) 
on a given dressing $b_{{\mathcal T}}\in \bar {\mathcal B}_{{\mathcal T}_{\ast}}$ is finite. 
Thus, the scope of nontrivial zero holonomy conditions (\ref{prod1}) drastically reduces and 
the task of ensuring consistency of translations of the population activity vectors over 
$\mathcal{T}_{\ast}$ becomes tractable. Nevertheless, zero curvature conditions (\ref{curv}) 
are in general quite restrictive and impose nontrivial constraints on the synaptic architecture 
of the place cell assemblies. As the simplest illustration, consider the case when the firing 
rates of all the place cells and readout neurons are the same: $f_{v_i}=f_{\sigma_{k}}=f$, 
and all the connection strengths from the place cells to the readout neuron in all cell assemblies 
are identical: $b_{\sigma_i,v_i}=\bar b$, giving a constant connection dressing 
$\bar b_{\mathcal{T}_{\ast}}$. It can be shown that in this case the resulting transfer matrix 
is idempotent, that is $M_{\hat\sigma}^2=\mathbf{1}$, so that the zero curvature condition 
(\ref{curv}) is satisfied identically for the even order elementary closed paths and cannot be 
satisfied if the paths' order is odd. Under more general and physiologically more plausible 
assumptions equation (\ref{curv}) does not necessarily restrict the order of the cell assemblies. 
However, the domain of permissible dressings, $\bar {\mathcal B}_{{\mathcal T}_{\ast}}$ is 
significantly restricted by (\ref{curv}), as compared to the domain occupied by the synaptic 
parameters of the unconstrained cell assembly networks. 

\section{Statistics of synaptic weights in the limit of weak synaptic noise}
\label{section:stat}

The zero curvature constraints (\ref{curv}) affect the net statistics of the synaptic 
weights. Since the structure of the full space of marginal dressings 
$\bar{{\mathcal B}}_{{\mathcal T}_{\ast}}$ and of the corresponding probability 
measures is too complex, we considered a family of connections parametrized as 
\begin{equation}
b_{\sigma,v}=\bar b\left(1+\varepsilon_{\sigma,v}+O(\varepsilon_{\sigma,v}^2)\right),
\label{param}
\end{equation}
in which the fluctuations $\varepsilon_{\sigma,v}$ are normally distributed,
\begin{equation}
P(\varepsilon_{\sigma,v}) =\frac{1}{\varepsilon \sqrt{\pi }}
e^{-\frac{\varepsilon_{\sigma,v}^2}{\varepsilon^2}}.
\label{gauss}
\end{equation}
In the absence of zero curvature constraints, cell assemblies are uncoupled and the 
synaptic fluctuations are statistically independent, so that the joint probability 
distribution of $\varepsilon_{\sigma,v}$ is
\begin{equation}
\mathcal{P}(\varepsilon)
=\prod_{\sigma,v}P\left(\varepsilon_{\sigma,v}\right).
\label{indep}
\end{equation}
Under zero-curvature conditions (\ref{curv}) the parameters of  the synaptic 
architecture are coupled (Figure~\ref{Coupling}) and the probability 
distribution for a particular variable $\varepsilon_{\sigma,v}$ is obtained by 
averaging the joint distribution (\ref{indep}) under delta-constraints:
\begin{equation}
\hat P(\varepsilon_{\sigma,v}) =C
\idotsint
\prod_{\hat\sigma}
\prod_{i=1}^{d+1}\delta \left(\kappa_{\hat\sigma,i}(\varepsilon)\right)
\mathcal{P}(\varepsilon)D'\varepsilon,
\label{eps}
\end{equation}
where $C$ is the normalization constant and 
$D'\varepsilon\equiv\prod'd\varepsilon_{\sigma,v}$ denotes integration 
over all $\varepsilon_{\sigma',v'}\neq \varepsilon_{\sigma,v}$.

In the Appendix it is demonstrated that for weak fluctuations, the shape of 
the distribution (\ref{gauss}) remains Gaussian, 
\begin{equation}
\hat P\left(\varepsilon_{\sigma,v}\right) =
\frac{1}{\sqrt{\pi}\varepsilon_{\ast}}e^{-\frac{\varepsilon_{\sigma,v}^2}{\varepsilon^2_{\ast}}}
\label{tune}
\end{equation}
but its width decreases: $\varepsilon_{\ast}<\varepsilon$.
Thus, zero curvature conditions narrow the distribution of the uncorrelated 
weights, i.e., produce a ``tuning" of the synaptic connections $b_{\sigma,v}$.
This result also applies to the synaptic weights: in cases where the place fields 
are distributed regularly, so that the coefficients $h_{\sigma,v_i}$ have a well 
defined mean, $\bar{h}$, and a small multiplicative variance, 
$$h_{\sigma,v_i}=\bar{h}\left( 1+\delta_{\sigma,v_i}\right),$$
$\delta_{\sigma,v_i}\ll 1$, the coefficients $\mu_{v_1,v_2}^{\sigma}$
are approximately defined by the ratios of the synaptic weights,
\begin{equation}
\mu_{v_{1},v_2}^{\sigma}=\frac{b_{\sigma,v_1}}{b_{\sigma,v_2}}=
\frac{w_{\sigma,v_1}}{w_{\sigma,v_2}}\frac{h_{\sigma,v_1}}{h_{\sigma,v_2}}
=\frac{w_{\sigma,v_1}}{w_{\sigma,v_2}}\left(1+\delta_{\sigma,v_1}-
\delta_{\sigma,v_2}\right) \approx \frac{w_{\sigma,v_1}}{w_{\sigma,v_2}},
\nonumber
\end{equation}
and therefore the zero curvature conditions produce the same effect on $w_{\sigma,v}$ 
as on $b_{\sigma,v}$, i.e., reduce the variability of synaptic weights. 

Understanding the effects produced by zero curvature constraints (\ref{curv}) on a wider range 
of fluctuations is mathematically more challenging. The qualitative results obtained here, however, 
may generalize beyond the limit of small multiplicative synaptic noise and could eventually be 
experimentally verified. 
A physiological implication of the result (\ref{tune}) is that the distribution of the unconstrained 
synaptic weights in a network that does not encode a representation of space (e.g., measured 
\emph{in vitro}) should be broader than the distribution measured \emph{in vivo} in healthy 
animals, which can be tested once such measurements become technically possible.

\section{Discussion}
\label{section:discussion}

The task of encoding a consistent map of the environment imposes a system of constraints
on the hippocampal network (i.e., on the coefficients $b_{\sigma,v}$) that enforce the 
correspondence between place cell activity and the animal's location in the physical world. 
Here we show that zero holonomy is a key condition, which is implemented by requiring that 
curvatures vanish at the pivot simplexes. This approach works within a combinatorial 
framework, but a similar intuition guided a geometric approach \cite{Issa}, where the place 
cells' ability to encode the location of the animal---but not the path leading to that location---was 
achieved by imposing the conditions of Stoke's theorem \cite{Dubrovnik} on the synaptic 
weights of the hippocampal network, which were viewed as functions of Cartesian coordinates. 
Our model is based on the same requirement of path-invariance of place cell population activity, 
implemented on a discrete representation of space---a dressed abstract simplicial complex 
$\mathcal{T}_{\ast}$---without involving geometric information about the animal's environment.

In particular, note that the concepts of ``curvature'' and ``holonomy'' are defined in 
combinatorial, not geometric, terms. This is an advantage in light of (and indeed was 
motivated by) recent work indicating that the hippocampus provides a topological framework 
for spatial experience rather than Cartesian map of the environment \cite{eLife}, and it also 
makes our model somewhat more realistic. 
It does, however, lead to a number of technical complications. For example, discrete 
connections (\ref{b}) defined over $\mathcal{T}_{\ast}$ are nonabelian \cite{Novikov}, 
so using the approach of \cite{Issa} would require a nontrivial generalization of Stoke's 
theorem, which is valid only in spaces with abelian differential-geometric connections 
\cite{Broda}. Our approach is based on the analysis of discrete holonomies suggested 
in the pioneering work of \cite{Novikov} which, in fact, explains the mathematical 
underpinning of the Stoke's theorem approach in both abelian and nonabelian cases 
\cite{Bredon,Dubrovnik,Sternberg}. Indeed, the zero-holonomy constraint ensures that 
no matter what direction the activity is propagated in the network (forward, backward, or 
skipping over some cell assemblies), the integrity of the spatial information remains intact.

{\bf Generality of the approach}.
A key instrument of our analyses is equation (\ref{Q}), which describes the conditions 
necessary for propagating spiking conditions over the cell assembly network. The exact 
form of this equation is not essential---a physiologically more detailed description of 
near-threshold neuronal spiking \cite{Poirazi1,Poirazi2,Wallach} could be used to establish 
more accurate zero holonomy and curvature constraints on the hippocampal network's 
synaptic architecture, which should be viewed as a general requirement for any spatial 
replay model. 

\begin{figure} 
\includegraphics[scale=0.7]{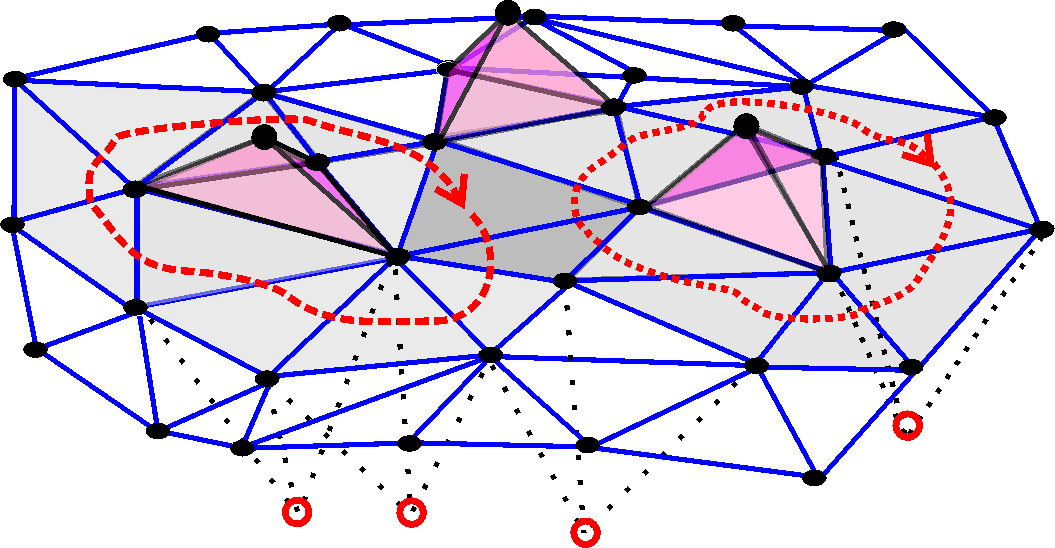}
\caption{\label{QManif} {\bf A replay in simplicial quasimanifold}. An example of a 
simplicial quasi-manifold containing $2D$ and $3D$ simplexes. The activity of cells in 
the $3D$ simplexes is induced from the $2D$ simplexes approaching its sides.
Two simplicial paths are shown by gray triangles, marked by red dotted lines.}
\end{figure}

The assumption of maximally-overlapping place cell assemblies may also be relaxed, 
since equation (\ref{Q}) can be applied in cases where the order of the cell assemblies 
varies, that is, when the simplicial complex $\mathcal{T}_{\ast}$ is not a manifold but 
a quasimanifold (see Figure~\ref{QManif} and \cite{Floriani,Lienhardt}). Unfortunately,
implementing the ``zero holonomy'' principle in this case would require rather arduous 
combinatorial analysis. For example, propagating the activity packets using (\ref{kk}) 
would impose relationships between the dimensionalities of the maximal simplexes and 
their placement in $\mathcal{T}_{\ast}$, i.e., require a particular cell assembly network
architecture.

{\bf Learning the constraints}. In this paper, the requirements (\ref{prod1}) and 
(\ref{curv}) enforcing path consistency of place cell replay are imposed on a 
fully trained network: it is assumed that the place fields have had time to stabilize 
\cite{Frank} and that the cell assemblies with constant weights $w_{\sigma,v_i}$ 
have had time to form \cite{Best}. 
In a more realistic approach, these constraints should modulate the hippocampal 
network's training process. For example, if the unconstrained network is trained by 
minimizing a certain cost functional $S(b_{{\mathcal T}_{\ast}})$ 
\cite{Hopfield,Hopfield1} then the constraints (\ref{curv}) would contribute an additional 
``curvature term" $R(b_{\mathcal{T}_{\ast}})$,
\begin{equation}
S(b_{\mathcal{T}_{\ast}}) \to S(b_{\mathcal{T}_{\ast}})+R(b_{\mathcal{T}_{\ast}}), 
\label{action}
\end{equation}
defined, e.g., via Lagrange multipliers $r_i$,
\begin{equation}
R(b_{\mathcal{T}_{\ast}})=\sum_{\hat\sigma\in \mathcal{T}_{\ast}} 
R_{\hat \sigma}(b_{\mathcal{T}_{\ast}})=\sum_{\hat \sigma,i}r_i\,\kappa_{\hat\sigma,i}.
\label{R}
\end{equation}

Physiologically, the network may be trained by ``ringing out'' the violations of the conditions 
(\ref{curv}) in the neuronal circuit, i.e., by replaying sequences and adapting the synaptic
 weights to get rid of the centers of non-vanishing holonomy.
Curiously, the role played by $R(b_{\mathcal{T}_{\ast}})$ in (\ref{action}) resembles the role 
played by the curvature term in the Hilbert--Einstein action of General Relativity Theory 
\cite{Dubrovnik}, which ensures that, in the absence of gravitational field sources, the solution 
of the Hilbert--Einstein equations describes a flat space-time. By analogy, the constraints imposed 
by (\ref{curv}) may be viewed as conditions that enforce ``synaptic flatness'' of the hippocampal 
cognitive map.

It is worth noting that the mechanism suggested here is an implementation of the zero holonomy 
condition in this simplest case of the reader-centric cell assembly theory that is consistent with 
physiology. The place cell readout might involve, instead of a single neuron, a small network of a 
few neurons (not yet identified experimentally), which  might require a different implementation 
of zero holonomy principle, depending on the specific architecture of such a network. If the readout 
network is a cluster of synchronously activated downstream neurons, then this cluster of cells could 
be viewed as a ``meta-neuron'' and the proposed approach would apply to this case as well. More 
complicated architectures would require modifications, but it is reasonable that the reproducibility of 
the population vector would require zero holonomy in all cases.

\section{Acknowledgments}
\label{section:acknow}

I thank V. Brandt and R. Phenix for their critical reading of the manuscript and the reviewers for 
helpful comments..

The work was supported in part by Houston Bioinformatics Endowment Fund, the W. M. 
Keck Foundation grant for pioneering research and by the NSF 1422438 grant.

\section{Appendix}
\label{section:appx}

{\bf Transfer matrix} construction is carried out for the $2D$ case, since higher 
dimensions are similar. In the matrix form, equation (\ref{v2sol1}) defined over the 
simplex $\sigma_p$, can be written as $\mathbf{f}_{q}=\hat M_{q,p}^{v_k v_j}\mathbf{f}_{p}$,
in which the matrix  
\begin{equation}
\hat M_{q,p}^{v_k v_j}=\left( 
\begin{array}{ccc}
1 & 0 & 0 \\ 
0 & 1 & 0 \\ 
\mu_{\sigma_p,v_k}^{\sigma_{p}} & -\mu_{v_i,v_k}^{\sigma_p} & 
-\mu_{v_j,v_k}^{\sigma_p}
\end{array}
\right),
\nonumber
\end{equation}
transfers the activity vector
\begin{equation}
\mathbf{f}_{p}^{\top }=\left( f_{\sigma_p},f_{v_i},f_{v_j}\right),
\nonumber
\end{equation}
defined over the incoming edge $[v_i,v_j]$ of the $p$-th simplex into the activity 
vector at the outgoing edge $[v_i,v_k]$ of the \emph{same} simplex (e.g., from the 
edge $[v_0,v_1]$ to the edge $[v_1,v_2]$ of $\sigma_1$ on Figure~\ref{ThickPath}A),
\begin{equation}
\mathbf{f}_{p}^{\top}=\left( f_{\sigma_p},f_{v_i},f_{v_{k}}\right).
\label{fp1}
\end{equation}
To ignite the readout neuron of the next cell assembly $\sigma_q$, which shares the 
edge $[v_i,v_k]$ with $\sigma_{p}$ the vector (\ref{fp1}) needs to be transformed into 
\begin{equation}
\mathbf{f}_{q}^{\top}=\left(f_{\sigma_q},f_{v_i},f_{v_{k}}\right)
\nonumber
\end{equation}
by the diagonal matrix $D_{p,q}={\rm diag}(f_{\sigma_q}/f_{\sigma_p}, 1,1)$.
Together, these two operations produce the transfer matrix 
\begin{equation}
M_{q,p}^{v_k v_j}=D_{p,q}\hat M_{q,p}^{v_k v_j}=\left( 
\begin{array}{ccc}
f_{\sigma_q}/f_{\sigma_p} & 0 & 0 \\ 
0 & 1 & 0 \\ 
\mu_{\sigma_p,v_k}^{\sigma_{p}} & -\mu_{v_i,v_k}^{\sigma_p} & 
-\mu_{v_j,v_k}^{\sigma_p}
\end{array}
\right).
\nonumber
\end{equation}
A direct verification shows that a product of $n$ transfer matrices that stat and 
end at the same simplex, has the form (\ref{MGamma}), in which $\kappa_{\Gamma,i}$
are $n$th order polynomials of the coefficients (\ref{mu}).

\begin{figure} 
\includegraphics[scale=0.7]{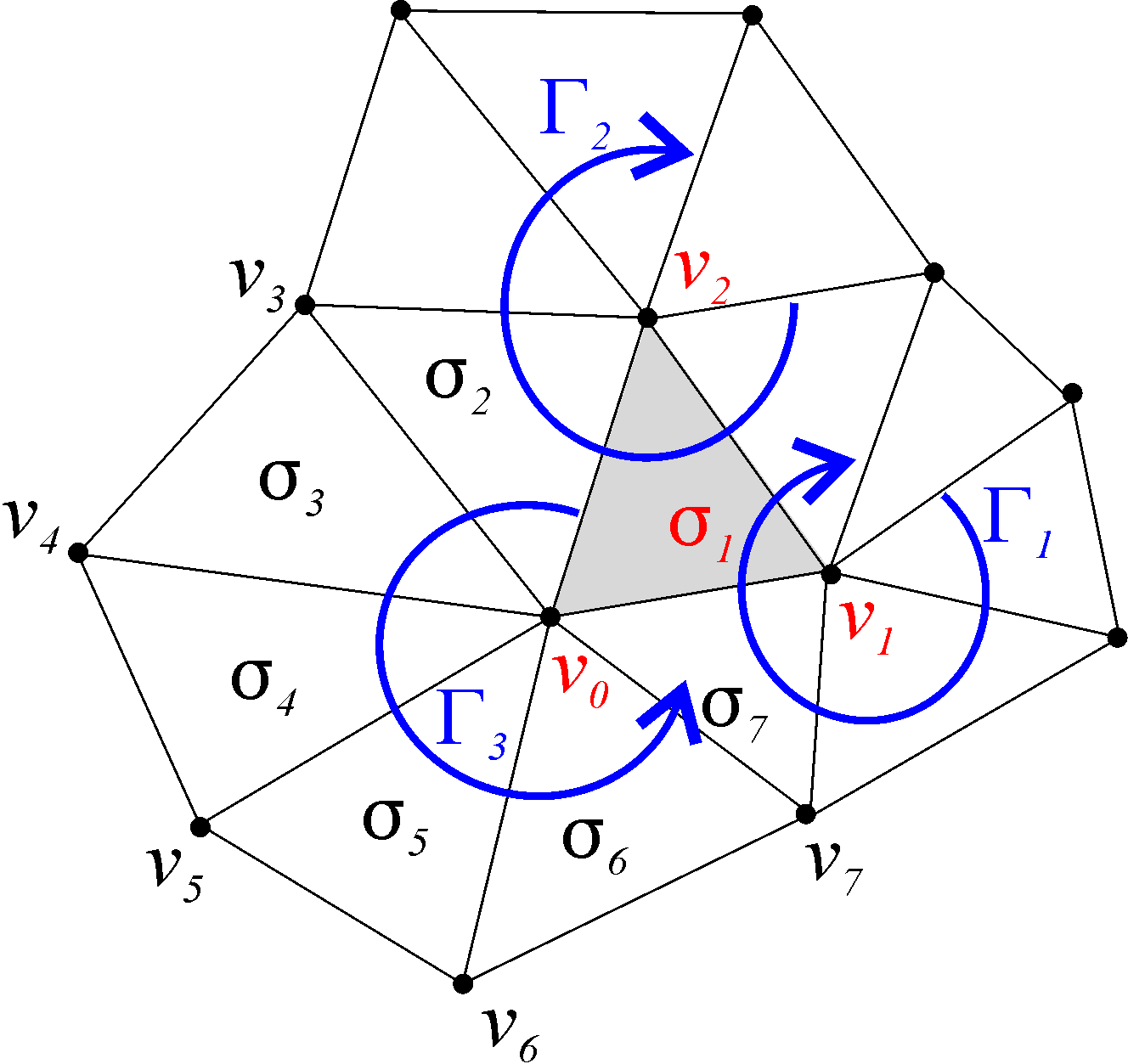}
\caption{\label{Coupling} {\bf Coupling between simplexes}. A schematic
illustration of a maximal simplex $\sigma_1$ span by three pivot vertexes, $v_0$, 
$v_1$ and $v_2$, and shared by three overlapping elementary paths $\Gamma_1$, 
$\Gamma_2$ and $\Gamma_3$ in a $2D$ cell assembly complex. As a result, the 
synaptic connectivity coefficients $b_{\sigma_1,v}$ will appear in three sets of discrete
curvatures, $\kappa_{v_0,i}$, $\kappa_{v_1,i}$ and $\kappa_{v_2,i}$, $i=1$, $2$, $3$, 
bootstrapping the constraints (\ref{curv}).} 
\end{figure} 

{\bf Tuning of the fluctuation distribution}. In case when the fluctuations are small,
$\varepsilon\ll 1$, the constraints (\ref{curv}) uncouple (Figure~\ref{Coupling})
yielding linearized curvature coefficients 
\begin{equation}
\kappa_{\hat\sigma,i}=\sum_{v\in \Gamma_{\hat\sigma}}
Q_{\hat\sigma,i;\sigma,v}\,\varepsilon_{\sigma,v},
\label{lincurv}
\end{equation}
where $Q_{\hat\sigma,i;\sigma,v}$ are constant coefficients and the summation is over 
the vertexes of the elementary path $\Gamma_{\hat\sigma}$ (Figure~\ref{Holonomy}B).
To simplify the expression (\ref{eps}), we rewrite it using indexes $p=(\sigma,v)$ and 
$l=(\hat \sigma,i)$, 
\begin{equation}
\hat P(\varepsilon_p) = C\idotsint
\prod_{l}\delta(\kappa_{l}(\varepsilon_{p'}))\mathcal{P}(\varepsilon_{p'})
\prod_{p'\neq p} d\varepsilon_{p'},
\nonumber
\end{equation}
and exponentiate the delta-functions,
\begin{eqnarray}
\hat P(\varepsilon_p) =C
\idotsint
\idotsint\limits_{\eta=-\infty}^{\infty}
\mathcal{P}(\varepsilon_{p'})
\prod_{l}d\eta_l e^{i\eta_{l}\kappa_{l}}
\prod_{p'\neq p} d\varepsilon_{p'}.
\label{expp}
\end{eqnarray}
Using the joint distribution (\ref{indep}) 
and the linearized expressions (\ref{lincurv}) in (\ref{expp}) produces
\begin{eqnarray*}
\hat P(\varepsilon_p)
&=&Ce^{-\frac{\varepsilon_{p}^2}{\varepsilon^2}}
\idotsint\limits_{\eta=-\infty}^{\infty}
\prod_{l}d\eta_{l}
e^{i\eta_{l}V_{lp}\varepsilon_{p}}
\idotsint\limits_{U_{\varepsilon }(\bar b_{\mathcal{T}_{\ast}})}
\prod_{p'\neq p} d\varepsilon_{p'}\,
e^{-\frac{\varepsilon_{p'}^{2}}{\varepsilon^2}} 
e^{i\eta_{l}V_{lp'}\varepsilon_{p'}},
\end{eqnarray*}
where the $V_{lp}$ are the coefficients obtained by collecting the 
terms proportional to $\varepsilon_{p'}$s produced by (\ref{lincurv}).
Completing the square and integrating over $\varepsilon_{p'}$ yields a 
Gaussian integral over a positive quadratic form $A=VV^{\top}$,
\begin{equation}
\hat P(\varepsilon_{p}) =
e^{-\frac{\varepsilon_{p}^2}{\varepsilon^2}}
\idotsint\limits_{\eta=-\infty}^{\infty}\prod_{l}d\eta_{l}
e^{i\varepsilon_{p}v_{p}\eta}e^{-\frac{1}{4}\varepsilon^2(\eta A\eta)}.
\label{PP}
\end{equation}
where $v_{p}$ is the $p$th row of the matrix $V$.
Evaluating (\ref{PP}) yields
\begin{equation}
\hat P(\varepsilon_{p})=
Ce^{-\frac{\varepsilon_{p}^2}{\varepsilon^2}-\frac{\varepsilon_{p}^2}{\varepsilon^2}v_p A^{-1}v_p}
=Ce^{-\frac{\varepsilon_{p}^2}{\varepsilon^2_{\ast}}}
\label{PP1}
\end{equation}
where
\begin{equation}
\varepsilon^2_{\ast}=\varepsilon^2\left(1+v_p A^{-1}v_p\right)^{-1}.
\label{epss}
\end{equation}
Since the second term in the parentheses is positive, $\varepsilon_{\ast}<\varepsilon$, which 
indicates narrowing of the uncoupled distribution (\ref{gauss}). 
The magnitude of the correction in (\ref{epss}) depends on the topological structure of the coactivity 
complex (e.g., its dimensionality $d$ and the statistics of the pivots' orders, $n$) and on the dressing 
parameters, $\bar {\mathcal B}_{{\mathcal T}_{\ast}}$. 
In the approximation (\ref{param}), $\varepsilon\ll 1$, the diaginal matrix elemens of the 
matrix $A$ are of the order $d \bar n$, and hence the $v_p A^{-1}v_p \sim d/\bar n$.

\newpage

\section{Citations}

\end{document}